\pdfoutput=1
\documentclass[%
 reprint,
nofootinbib,
 amsmath,amssymb,
 aps, prd,
floatfix,
showkeys,
]{revtex4-1}

\usepackage{graphicx}
\usepackage{dcolumn}
\usepackage{bm}
\usepackage{amsmath}
\usepackage{slashed}

\begin{document}

\preprint{}

\title{Quantization of ratio gravity in Minkowski spacetime and mass generation mechanism}

\author{Jackie C.H. Liu}
 \email{chjliu@ust.hk}%

\affiliation{%
Department of Physics, The Hong Kong University of Science and Technology, Clear Water Bay, Kowloon, Hong Kong, P.R.China
}%

\date{\today}

\begin{abstract}
Recently, the theory of ratio gravity (RG) has introduced a new description of spacetime curvature and gravity as well as dark energy.  This paper proposes a quantization approach through a simple model for RG.  By solving the ratio gravity equations in Minkowski spacetime, we found there are two wave solutions coupled to a symmetry broken scalar field.  We postulate the related Lagrangian with massless fermion doublets coupling a scalar field, and perform the related one-loop calculation.  The theory provides the non-zero and positive masses for three generations of three types of massive fermions: apparently as leptons (electron, muon, tau), and quarks (up, charm, top, down, strange, bottom).  The predicted masses of the top and bottom quarks (181 GeV and 3.5 GeV) are close to the experimental data.  We address the possibility that RG quantization may offer mass generation mechanism for fermions.
\end{abstract}

\maketitle

\section{\label{sec:level1}INTRODUCTION}
 
A theory called ratio gravity (RG) \cite{LiuWang2018}\cite{Liu2016} proposed that the curvature of 3+1 spacetime originates from a deformation of a cross ratio, where a similar mathematical structure to general relativity emerges.  

In the context of cosmology, the theory predicts a component of dynamical dark energy.  Recent observations show some evidence that dark energy may have similar dynamics \cite{Zhao:2017cud}. An observational best-fit polynomial plot for $w(z)$, with an interesting $w(z)=-1$ crossing,%
\footnote{The crossing of $w(z)$ to the line of $w=-1$ has been studied in the literature of dynamical dark energy. See \cite{Feng:2004ad} and \cite{Copeland:2006wr, Cai:2009zp, Bamba2012, Wang:2016och} for reviews.}, is compared to the plot of $w(z)$ of RG, with good agreement.  The RG theory provides an alternative explanation to the expansion history, matching well with observation.  (We refer interested readers to a recent and more extensive discussion in \cite{LiuWang2018}).  Yet the theory so far lacks quantization. 

In this paper, we propose the principle of the quantization of RG theory: identifying the solution of RG theory for given spacetime geometry, postulating the related Lagrangian as a model of quantum fields, and finally looking for a renormalization theme for the quantum observables.

We follow the principle above while working out a simple model.  This paper consists of three parts.  Part I (Section 2): the Minkowski solution from RG theory; part II (Sections 3 and 4): constructing a potential renormalizable Lagrangian through second quantization \cite{Negele1988}; and part III (Section 5): the renormalization scheme and mass generation as well as prediction of certain quark masses.  The complete renormalizable analysis is not covered in this paper; therefore, we refer the term renormalizable to the limited condition, i.e., up to the one-loop calculation.

Some supersymmetry theories and beyond-standard-model theories such as \cite{Dobrescu2010} suggest that certain fermion masses are generated by one-loop induced couplings, the mass generation mechanism in this paper is also caused by one-loop amplitude, though the detail is different.

We found the theory requires the doublet structure of quantum fields to realize the Minkowski solution from RG theory, and the $U1\otimes SU2$ gauges ($SU2$ and $SU2$ sub-algebra of $SU3$) in order to construct the renormalizable model.  The model implies there are different vacuum expectation values (VEVs) for fermions, similar to the two-Higgs-doublet model (2HDM) \cite{Dobrescu2010}.

As a result, the theory derives a positive mass formula, $M_{ab}=f_a^2 g_b^2 \Lambda_{fer}(1-2 c_a g_b)^2 e^{2 v_a g_b}$ (in Section 5), for all nine fermions (nine out of 12 fermions in the standard model, except the neutrino family).  They are apparently different representations of the same wave solution of the Minkowski solution from RG theory.  We probe the quark parameters by four fermion masses (up, down, charm, and strange).  Then we calculate the masses of the top and bottom quarks (181 GeV and 3.5 GeV), which is close to the current data [FIG.~\ref{fig:masses}].  We also probe the lepton parameters by electron, muon, and tau masses to check the consistency of the theory.  The current model does not include the weak interactions and higher-loop calculations, the corrections to the mass formula is currently unknown.

\section{SOLVING THE RATIO GRAVITY EQUATION IN Minkowski spacetime}

By solving the equations of RG theory\textendash Y equation (\ref{eq:YEqn}), Gal-D-commuting relation (\ref{eq:GalDComEqn}), and RG Bianchi equations (\ref{eq:DJDM0}),
\begin{equation}
D_{ab\prime }Y=B_{ab\prime }Y,							\label{eq:YEqn}
\end{equation}
\begin{multline}
\hat{J}_{ac}(Y)=J_{ac}Y, \hat{M}_{a\prime c\prime }(Y)=M_{a\prime c\prime }Y, \\ 
D_{bd'}\hat{J}_{ac}Y=\hat{J}_{ac}D_{bd'}Y, D_{bd'}\hat{M}_{a'c'}Y=\hat{M}_{a'c'}D_{bd'}Y  \label{eq:GalDComEqn}, 	
\end{multline}
\begin{equation}
	D^{b}_{d\prime}J_{ab}(Y)=D_{a}^{c\prime}M_{c\prime d\prime}(Y)	\label{eq:DJDM0}
\end{equation}
respectively, Bianchi constraints, and RG-NP equations\footnote{Both Bianchi constraints and RG-NP equations are not shown because they are solved trivially in this paper.  For details, we refer readers to \cite{LiuWang2018}.} in Minkowski spacetime, we set $Be_{ab'} = 0, Bh_{ab'} = 0, Bf_{ab'} \neq 0$\footnote{$Bf_{ab'}, Be_{ab'}$, and $Bh_{ab'}$ denote the f-component, e-component, and h-component of the $B_{ab'}$ matrix, where the f-component, e-component, and h-component are in $sl_2$ algebra (f,e,h) respectively.}, and the scalar part of $B_{ab'} \neq 0$, so $J_{ab},M_{c'd'} = 0$.  So the Bianchi constraints and RG-NP equations are trivially solved with zero spin coefficients.  The dyad differential operators are simply $\partial_{a,b'}=\sigma_{a,b'}^\mu \partial_\mu$, where $\sigma_{a,b'}^\mu$ are Pauli matrices.

Even thought we are working on a simple case of Minkowski spacetime, RG equations (\ref{eq:GalDComEqn}) and (\ref{eq:DJDM0}) have specific differential equations to be solved because of the definitions of curvature operators $\hat{J}_{ac}$ and $\hat{M}_{a\prime c\prime }$\cite{LiuWang2018}.  Thus, the solution must respect a certain symmetry because of the nature of the overdetermined system of differential equations generally in RG theory.

There are two sets of solutions found (partial differential equations): $y_h, y_f$ solutions to denote the h-component and f-component of Y.  The solutions are
\begin{equation}
\partial_{\mu}y_h=-i g \frac{k^{\mu}}{k^0} (c_1 + \Phi f_1)y_h,
\end{equation}
\begin{equation}
\partial_{\mu}y_f=-i g \frac{k^{\mu}}{k^0} (c_2 + \Phi f_2)y_f, 
\end{equation}
where $\Phi$ is a trival wave function with momentum $k^{\mu}$, and $f_1,f_2, g, c_1$, and $c_2$ are parameters of the solutions.

The wave solution is found for each case, so we construct the $y$ \textit{wave} (as doublet) coupled to the symmetry-breaking scalar wave which is in the form $c_a + f_a \Phi$.  Our model relates different y-component solutions with same $\Phi$ field (i.e., relates two solutions in solutions space).  Physically, it means we model a single scalar field to couple the fermion doublet.

We look for the fermion field equation, so in order to get a coupled spinor equation for quantization, we tensor an arbitrary spinor $\xi$ to $y$:
\begin{equation}
	i\slashed{\partial} \xi\otimes y=g\frac{ \slashed{k} \hat{\Phi } \xi\otimes y}{k^0}
	\label{eq:ywaveEoM},
\end{equation}
where $\hat{\Phi }:=
\left(
\begin{array}{cc}
 c_1+\Phi  f_1 & 0 \\
 0 & c_2+\Phi  f_2 \\
\end{array}
\right)$.

\section{QUANTUM FIELDS AND R-GAUGE}

In the previous section, we construct a wave equation (\ref{eq:ywaveEoM}) in the doublet form of the $y$ solution by both $y_h$ and $y_f$ solutions.  In this section, we introduce a gauge transformation for the R-spinor: R-gauge.  The purpose is to create a nilpotent matrix structure in order to construct a renormalizable model in the next section.

Throughtout the paper, we use the spinor form in L, R components as $\left(\begin{array}{c} L \\ R \\ \end{array}\right)$.

For the R-spinor, we use $U(1)\otimes SU(2)$ gauge transformation to wave equation (\ref{eq:ywaveEoM}) in order to create a nilpotent matrix structure $\eta(\hat{\Phi'}):=\left(
\begin{array}{cccc}
 0 & 0 & \hat{\Phi' } & 0 \\
 0 & 0 & 0 & \hat{\Phi' } \\
 0 & 0 & 0 & 0 \\
 0 & 0 & 0 & 0 \\
\end{array}
\right)$, where $\hat{\Phi'}$ is the R-gauge generated symmetry-breaking scalar field.

Since we need the doublet for the R-gauge transformation to create an $\eta$ nilpotent matrix in order to have a potential renormalizable model in the next section, SU2 is used as local gauge for the R-gauge.

There are two natural choices of SU2: the first is the ordinary SU2 (called lepton-R-gauge), and the second (called quark-R-gauge) is SU2 sub-algebra in SU3.  Once we have done the R-gauge, the transformed wave equation reads as
\begin{equation}
	i\slashed{\partial}\Psi=g(Id_4\otimes \hat{\Phi } - \eta(\hat{\Phi'}) ) \Psi
	\label{eq:gaugedywaveEoM},
\end{equation} where $\Psi$ is R-gauged $\xi\otimes y$, and we require that $\xi$ is an eigen-state of $\slashed{k}$ with mass $m_{\Phi}$.
Note, we also use the condition that k is time-component-dominated and $\partial_{t}U \gg \partial_{x}U, \partial_{y}U,\partial_{z}U$.  Other models for the R-gauge are possible, but we do not cover them in this paper.

For the lepton-R-gauge, $\hat{\Phi'}$ tranforms as $\hat{\Phi'_{l}} \to \left(
\begin{array}{cc}
 0 & 0 \\
 0 &  (\text{$\mathit{c}$1}_l-\text{$\mathit{c}$2}_l+\Phi  f_l) \\
\end{array}
\right)$.  The parameters are redefined to relate to the lepton-R-gauge.  Only the difference of the diagonal components can be kept after transformation, and we use gauge freedom to preserve the diagonalised $\hat{\Phi'}$ (for preserving the commuting property of both terms  of the rhs of wave equation (\ref{eq:gaugedywaveEoM}), which is needed in the next section).

For the quark-R-gauge,\footnote{Ysing $\lambda_3,\lambda_4,\lambda_5$, and $\lambda_8$ of the Gell-Mann matrices.} we found two cases (called A-type and B-type):
$\hat{\Phi'_{A}} \to \left(
\begin{array}{ccc}
 0 & 0 & 0 \\
 0 &  \left(\text{$\mathit{c}$1}_q+\Phi  \mathit{f}_{q1}\right) & 0 \\
 0 & 0 & 0 \\
\end{array}
\right)$, $\hat{\Phi'_{B}} \to \left(
\begin{array}{ccc}
  \left(\text{$\mathit{c}$2}_q+\Phi  \mathit{f}_{q2}\right) & 0 & 0 \\
 0 & 0 & 0 \\
 0 & 0 & 0 \\
\end{array}
\right)$.  The parameters are redefined to relate to the quark-R-gauge, and note that the lower right corner position of $\hat{\Phi'_{A}},\hat{\Phi'_{B}}$ must be zero because we use the first and second rows for the y doublet.  Only one component of the diagonal components can be kept, so each case has one surviving component.  This is similar to the 2HDM model \cite{Dobrescu2010} in which there are two vacuum expectation values for up and down quarks.

\section{THE LAGRANGIAN}

In the context of inflation-cosmology, Wang explained the spectator field \cite{Parker1969} and how it is quantized in de Sitter space \cite{Wang2014}.  The spectator field $\sigma$ can be quantized as:
$\sigma=v_k(t) a_k + v_k(t)^* \overset{\dagger}{a_{-k}}$, where the mode function  $v_k(t)$ satisfies the equation of motion of $\sigma$.  In this section, we postulate the Lagrangian by first finding the related field operators of the R-gauge wave equation.  Similarly, we introduce the $\Psi$ field as analog to $\sigma$, and the related mode function also satisfies the equation of motion of $\Psi$. 

By writing $\Psi$ as follow, 
\begin{equation}
\Psi =\frac{\int \mathbf{d}^3p \left(\overset{s}{\mathbf{a(x)}_p} e^{-i px} \overset{s}{u_p}+e^{i px} {\overset{s}{\mathbf{b(x)}_p}}^\dagger \overset{s}{v_p}\right)}{8 \sqrt{2} \pi ^3 \sqrt{\mathcal{E}_p}},
\label{eq:PsiQWaveForm}
\end{equation}
where u and v are massless basis spinors, and $a(x)$ and $b(x)$ are unknown operators.  Substituting (\ref{eq:PsiQWaveForm}) to the wave equation (\ref{eq:gaugedywaveEoM}), we have\footnote{Use of commutativity property of scalar field operator.}
\begin{equation}
 \overset{s}{a(x)_p}=e^{\frac{g (Id_{4}\otimes\hat{\Phi } -\eta(\hat{\Phi'}) )}{m_{\Phi}}}\overset{s}{a_{0}}_{p} ,
\end{equation}
\begin{equation}
 \overset{\dagger }{\overset{s}{b(x)_p}}=e^{\frac{g (Id_{4}\otimes\hat{\Phi } -\eta(\hat{\Phi'}) )}{m_{\Phi}}}\overset{\dagger }{\overset{s}{b_0}_p},
\end{equation} where $\overset{s}{a_{0}}_{p}$ and $\overset{\dagger }{\overset{s}{b_0}_p}$ are ordinary annhillation and creation operators for the massless fermion.

According to the kinetic-energy-operator in \cite{Negele1988}, we postulate a modified kinetic-energy-operator as follows in a non-relativistic limit: 
\begin{multline}
\int d^3p\frac{\overset{{}^{\wedge}}{p}^2}{2 m}\psi^\dagger  \psi \to 
\\  \int d^3p\frac{\overset{{}^{\wedge}}{p}^2}{2 m} \psi^\dagger(e^{\frac{g (Id_{4}\otimes\hat{\Phi } -\eta(\hat{\Phi'}) )}{m_{\Phi}}})^\dagger e^{\frac{g (Id_{4}\otimes\hat{\Phi } -\eta(\hat{\Phi'}) )}{m_{\Phi}}}\psi, 
\end{multline}
where $\psi$ is the massless fermion field operator.  Note that because of the factor to fermion field and its adjoint, the proposed relativistic Lagrangian of the massless fermion is
\begin{equation}
\mathcal{L}_{fer} = \bar{\psi }  (e^{\frac{g (Id_{4}\otimes\hat{\Phi } -\eta(\hat{\Phi'}) )}{m_{\Phi}}})^\dagger e^{\frac{g (Id_{4}\otimes\hat{\Phi } -\eta(\hat{\Phi'}) )}{m_{\Phi}}} (i\slashed{\partial})\psi.
\end{equation} We expand the exponential terms in the Lagrangian,  and make use of the approximation that the scalar field is a small fluctuation compared to the vacuum energies of $\hat{\Phi}$ (e.g., $ f_1 \Phi \ll c_1$).  We further make use of the property of $\eta(\hat{\Phi'})$ that terminates the expansion (i.e., the nilpotent algebraic property), and recalling the time-component-dominated property of the scalar field, we get\footnote{Interaction terms with field derivative changed to momentum.}
\begin{multline}
\mathcal{L}_{fer}\simeq -\left(
\begin{array}{cc}
 R^{\dagger } & L^{\dagger } \\
\end{array}
\right)  e^{\frac{2 gV}{m_{\Phi}}} (
\begin{array}{c}
 \frac{g \hat{\Phi'} \overline{\sigma}^\nu p_\nu L}{m_{\Phi}} \\
 \frac{g \hat{\Phi'} \sigma^\nu p_\nu R }{m_{\Phi}} \\
\end{array}
) \\ +
\frac{g^2  L^{\dagger } e^{\frac{2 gV}{m_{\Phi}}} \hat{\Phi'}  \hat{\Phi'} \overline{\sigma}^\nu p_\nu L}{m_{\Phi}^2}\\
+(\begin{array}{cc} R^{\dagger } & L^{\dagger } \\ \end{array})e^{\frac{2 gV}{m_{\Phi}}}i \slashed{\partial}
 (\begin{array}{c} L \\ R \\
\end{array}),
\end{multline}
where $L$ and $R$ are the left and right massless fermion respectively, $V$ is the doublet form of VEV in the exponent: $\hat{\Phi}_{\Phi \to 0}$.   It is worth stating again that without the nilpotent property of $\eta(\hat{\Phi'})$, the Lagrangian cannot be renormalized because of the high power terms; this is why we needed to introduce the R-gauge in the previous section.  The doublet nature of the y solutions and $SU(2)$ are closely related to constructing a potential renormalizable field theory.

After introducing $\gamma^5$, we read off the Feynman rules for Yukawa interactions and the pseudo interactions for 3-point-interactions (i.e., $\bar{\psi } \Phi  \psi$),
\begin{equation}
\frac{i g \mathit{f}_a \left(g \mathit{c}_a-m_{\Phi}\right) e^{\frac{2 g v_a}{m_{\Phi}}}}{m_{\Phi}^2} \slashed{p},
\frac{i g^2 \mathit{c}_a \mathit{f}_a \gamma^5 e^{\frac{2 g v_a}{m_{\Phi}}}}{m_{\Phi}^2} \slashed{p},
\end{equation} and the vertices for 4-point-interactions (i.e., $\bar{\psi } \Phi \Phi \psi$),
\begin{equation}
\frac{i g^2 \mathit{f}_a^2 e^{\frac{2 g v_a}{m_{\Phi}}}}{2 m_{\Phi}^2} \slashed{p},
\frac{i \gamma ^5 g^2 \mathit{f}_a^2 e^{\frac{2 g v_a}{m_{\Phi}}}}{2 m_{\Phi}^2} \slashed{p},
\end{equation}
where $f_a, c_a$, and $v_a$ are the corresponding parameters ($v$ is the VEV in the doublet $V$) for the lepton, quark type-A, and quark type-B indexed by $a$.  In particular, $c_l$ is $c1_l-c2_l$.

The propagators for the fermion field and scalar field are $i e^{-2 \frac{g v_a}{m_{\Phi}}} \frac{\slashed{p}}{p^2}$ and  $i\frac{1}{k^2-m_{\Phi}^2}$ respectively.  We use the ordinary symmetry breaking scalar field Lagrangian ($\Phi^2$-theory) to model the $\hat{\Phi'}$ field.

\section{MASS GENERATION}

We calculate the one-loop amplitude for Yukawa interaction (the external momentum of the scalar field is zero) and self-energy of the fermion.\footnote{We perform the one-loop calculation by the symbolic software FeynCalc 9.2.0\cite{Shtabovenko2016} .}  Note that we need the pseudo interaction vertices during the one-loop calculation; however, we only focus on the one-loop amplitude of the non-pseudo interactions because they are related to the mass generation for our model.  Finally, we make use of the MMS scheme as follows:
$\epsilon \to \frac{1}{-\log (\frac{\pi  \mu ^2}{m_{\Phi } {}^2})+\log \left(\mu _{\overline{\text{MS}}}^2\right)+\gamma -1+2 \log(\pi )}$.

Similar to \cite{Langacker1995}, we may put tree-level and one-loop-level contributions together.  First, we define the virtual counter term ($\Delta_1$) with the relationship to the one-loop amplitude of 3-point-interaction:
$ (i \text{$\Delta_1$} g )\slashed{p}=\text{amplitude}$, where
\begin{multline}
\text{$\Delta_1$}=\\
\pi ^2 g^2 \mathit{f}_a^3  \log(  \mu _{\overline{\text{MS}}}^2) \left(g \mathit{c}_a-m_{\Phi }\right) \left(2 g \mathit{c}_a \left(2 g \mathit{c}_a-m_{\Phi }\right)+m_{\Phi }^2\right) \\
e^{\frac{2 g v_a}{m_{\Phi }}}  /(m_{\Phi }^4). \label{eq:delta1}
\end{multline}
Since the massless fermion causes $\slashed{p}=m_{\epsilon}$ (defined as infinitesimal fermion mass) $ \to 0$, the amplitude can be finite as the infinity from the virtual counter term $\times$ $m_{\epsilon}$ can be finite.  We call this \textit{finite-renormalization}.  The self-energy of the fermion provides the virtual counter-term ($\Delta_2$): 
\begin{multline}
\Delta_2 = 
\frac{\pi ^2 g^2 \mathit{f}_a^2 \log ( \mu _{\overline{\text{MS}}}^2 ) \left(m_{\Phi }-2 g \mathit{c}_a\right){}^2 e^{\frac{2 g v_a}{m_{\Phi }}}}{2 m_{\Phi }^2}.  \label{eq:delta2}
\end{multline}
Unlike the ordinary renormalization method that adds a counter-term to subtract infinity from the loop amplitude, we harmlessly add infinitesimal mass $m_{\epsilon}$ to the interaction term.  To include both tree-level and one-loop-level contributions, we define the rescaling factors $Z_1=1+\Delta_1$, $Z_2=e^{2 \frac{g v_a}{m_{\Phi}}}+\Delta_2$.  The renormalized interaction term is $-g\bar{\psi}(Z_1 m_{\epsilon}) \psi$.  The renormalized Lagrangian is then
\begin{equation}
\mathcal{L}_{re} = \bar{\psi}(Z_2 i\slashed{\partial}) \psi-g\bar{\psi}(Z_1 m_{\epsilon}) \psi + \mathcal{L}_{\Phi'}.
\end{equation}

To recover the usual Dirac field as the effective Lagrangian, we have $Z_1=Z_2$.  Considering this condition, equation (\ref{eq:delta1}), and equation (\ref{eq:delta2}), implies cubic roots of the g-polynomial to get the effective massive Dirac field, and we set $m_{\Phi} \to 1$ as our convention:
\begin{equation}
2 f (c g-1) (2 c g (2 c g-1)+1)-(1-2 c g)^2=0. \label{eq:gPoly}
\end{equation}
Observable mass is defined as $M=Z_1 m_{\epsilon}$, and it can be expressed as:
\begin{equation}
	M_{ab}=f_a^2 g_b^2 \Lambda_{fer}(1-2 c_a g_b)^2 e^{2 v_a g_b}, \label{eq:mEqn}
\end{equation}where $\Lambda_{fer}=\frac{\pi ^2}{2}m_{\epsilon} \log \left(\mu _{\overline{MS}}^2\right)$, a and b indices denote different types of fermions (electron-type, up-quark-type, down-quark-type) and different roots of the g-polynomial as generation indices respectively.  The constant $\Lambda_{fer}$ sets the mass-scale constant for fermions, and the expression of mass $M$ provides three roots for each generation for two cases: lepton type and mixing of quark-A type and quark-B type for the electron's and quark's families respectively.  Since we do not cover the electromagnetic and weak interactions in this paper, we can only identify the quark favours by the mass values.

Finally, we re-define the fermion field and $\slashed{\partial}$ to get the normal form of the Dirac Lagrangian: 
\begin{equation}
\mathcal{L}_{norm} = \bar{\psi}_{ab}(i\slashed{\partial}-M_{ab})\psi_{ab}+ \mathcal{L}_{\Phi'}.
\end{equation}

\begin{figure}[ht]
\includegraphics[width=200pt]{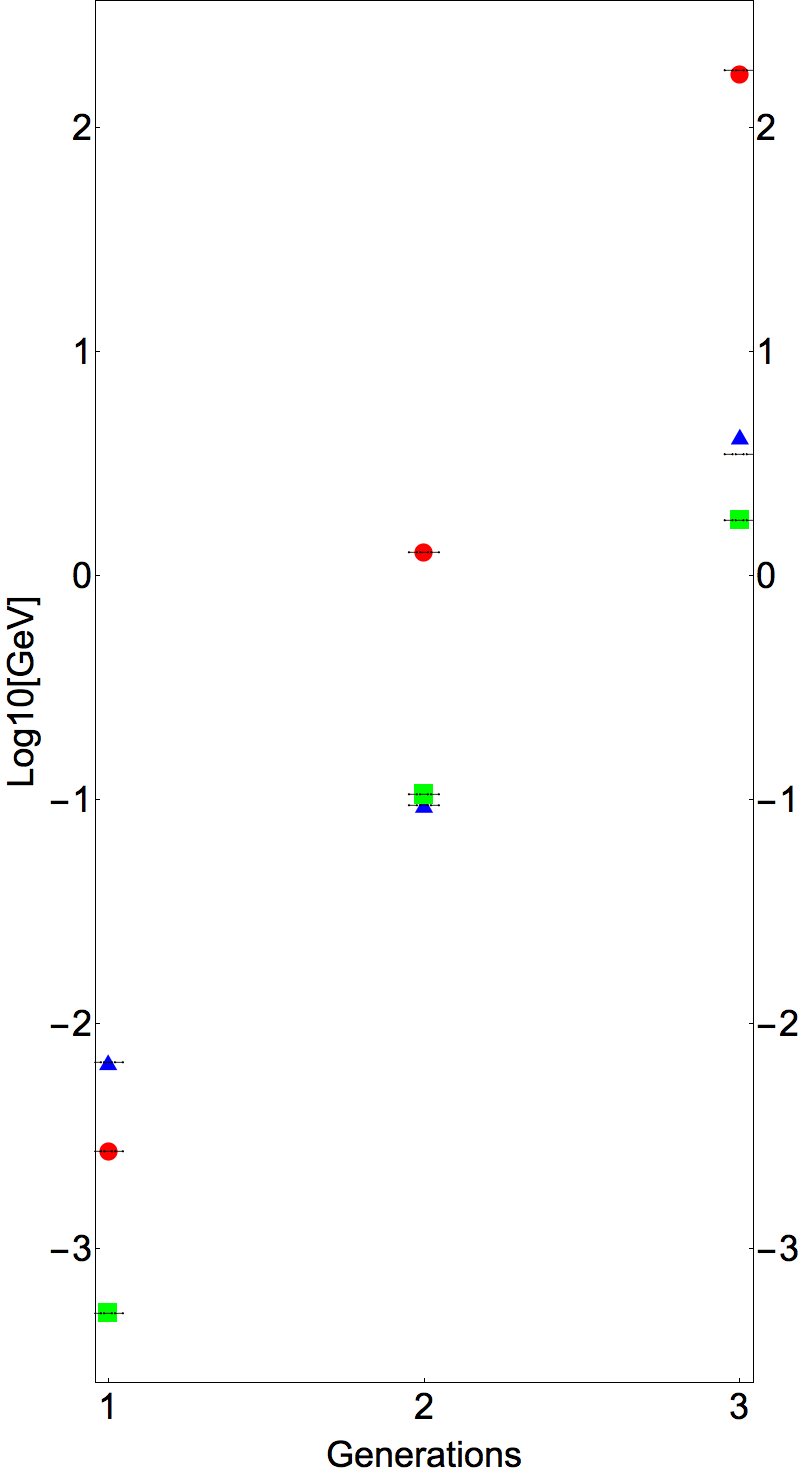}
\caption{\label{fig:masses} Fermion mass logarithm plot: electrons data (square), up quarks data (circle), down quarks data (triangle) \cite{Tanabashi2018}; the two top-right lines are the predicted top and bottom masses from mass formula (\ref{eq:mEqn}) 
with known experimental data: fermion masses including electron, muon, tau, and quarks (u, d, s, c).  Other lines are plotted to show the consistency of using the mass formula to match the experimental data to the parameters of the model.}
\end{figure}

To probe and check the consistency of the calculated masses versus the experiment data, we use mass data from particle group \cite{Tanabashi2018}.\footnote{Higgs mass=125.18 GeV, electron mass =0.512 MeV, muon mass=105.7 MeV, tau mass=1.777 GeV; quark masses for u, c, and t are 2.7 MeV, 1.275 GeV, and 173.1 GeV respectively; quark masses for d, s, and b are 6.75 MeV, 95 MeV, and 4.18 GeV respectively.}  The following values of parameters are found:
$\text{$\mathit{c}$2}_l\to 5.38738,\text{$\mathit{c}$1}_l\to 0.560435,\mathit{f}_l\to -0.492108,\text{$\mathit{c}$2}_q\to 0.77389,\text{$\mathit{c}$1}_q\to -2.69423,\mathit{f}_{\text{q1}}\to -0.489652,\mathit{f}_{\text{q2}}\to -0.50368,$ where we make use of the following data: fermion masses including electron, muon, tau, and quarks (u, d, s, c).  The predicted masses of the top quark and bottom quark are close to the current data (181 GeV and 3.5 GeV) [FIG.~\ref{fig:masses}].  Because of the lack of inclusion of the weak bosons contributions and higher-loop calculation, the corrections are not known yet.

The values of the parameters are not totally arbitrary, because the g-polynomial (\ref{eq:gPoly}) allows only certain $f$ values for real roots of $g$ and a given $c$ value.  For example, if 
$c$ is 1.39 (i.e., 174 GeV), $f$ is only allowed in the range $ -0.56 < f < 0$.

It is theoretically appealing that mass formula (\ref{eq:mEqn}) provides a positive mass solution in which the theory gives rise to a mass gap.

\section{Discussion}

In previous work \cite{LiuWang2018}\cite{Liu2016}, we introduced RG theory to describe the mathematical structure of gravity and apply it to the problem of dark energy.  In this paper, we develop the quantization principle and illustrate it by solving the equations of RG theory in the context of Minkowski spacetime, postulating the related Lagrangian, and calculating the mass formula.

We discover that the theory may offer a mass generation mechanism which apparently predicts the masses of the top and bottom quarks.  The theory also suggests that the lepton family and the quark families are originated by different SU2 algebras of gauge transformation of the cross ratio context $y$.  Interestingly, the generations of fermion may be explained by three distinct roots of cubic polynomial (\ref{eq:gPoly}).

We propose the future development of theory to include the weak interactions because the massive W and Z bosons should contribute to the mass formula.  The full renormalizability of the theory with more complete set of standard model interactions should also be covered next.
 
A complete quantization of a gravity theory must include the curved-spacetime geometry.  We leave the application to the curved-spacetime geometry by the principle of quantization stated in this paper to the future work.\\

\begin{acknowledgments}
JCHL would like to thank Professor Wang Yi for the valuable comments and advices.
This work is supported in part by ECS Grant 26300316 and GRF Grant 16301917 from the Research Grants Council of Hong Kong.
\end{acknowledgments}


\bibliography{QRG_1_0}

\end{document}